\documentclass[twocolumn,showpacs,preprintnumbers,amsmath,amssymb,floatfix]{revtex4}
\usepackage{graphicx}
\usepackage{amsfonts,latexsym}
\begin{document}
\title{
Energy level statistics for models of coupled single-mode Bose-Einstein
condensates
} 

\author{Silvio R. Dahmen}  
\email{dahmen@if.ufrgs.br} 
\affiliation{ Instituto de F\'{\i}sica da UFRGS, Caixa Postal
15051, Porto Alegre, 91501-970, Brasil} 
\author{Jon Links} 
\email{jrl@maths.uq.edu.au} 
\author{Ross H. McKenzie} 
\email{mckenzie@physics.uq.edu.au}
\author{Huan-Qiang Zhou} 
\email{hqz@maths.uq.edu.au} 
\affiliation{Centre for Mathematical Physics, The University of
Queensland, Brisbane, 4072, Australia}

\begin{abstract}
We study the distribution of energy level spacings in two models
describing coupled single-mode Bose-Einstein condensates. Both models
have a fixed number of degrees of freedom, which is small 
compared to the number of
interaction parameters, and is independent of the dimensionality of the
Hilbert space. We find that
the distribution follows a universal Poisson form independent of the
choice of coupling parameters, which is indicative of the
integrability of both models. These results complement those for
integrable lattice models where the number of degrees of freedom
increases with increasing dimensionality of the Hilbert space. 
Finally, we also show that for one model the inclusion of an additional
interaction which breaks the integrability leads to a non-Poisson
distribution.

\end{abstract}

\maketitle  
                     

\def\aa{\alpha} 
\def\bb{\beta}
\def\a{\hat a}
\def\b{\hat b}
\def\d{\dagger}
\def\de{\delta} 
\def\e{\epsilon}
\def\ve{\varepsilon}
\def\g{\gamma}
\def\K{\kappa}
\def\ap{\approx}
\def\l{\lambda}
\def\o{\omega}
\def\t{\tilde{\tau}}
\def\s{\sigma}
\def\D{\Delta}
\def\L{\Lambda}
\def\T{{\cal T}}
\def\TT{{\tilde{\cal T}}}
\def\E{{\cal E}} 
\def\f{\overline{f}}
\def\q{\overline{q}}
\def\tp{\otimes}
\def\I{\mathcal{I}} 
\def\N{\mathcal{N}}
\def\H{\mathcal{H}}
\def\rar{\rightarrow}

\def\bea{\begin{eqnarray}}
\def\eea{\end{eqnarray}}
\def\ba{\begin{array}}
\def\ea{\end{array}}
\def\no{\nonumber}
\def\le{\langle}
\def\re{\rangle}
\def\l{\left}
\def\r{\right}
\def\o{\omega}
\def\d{\dagger}
\def\nn{\nonumber}
\def\j{{ {\cal J}}}
\def\n{{\hat n}}
\def\A{{\cal A}}
\def\TT{{\tilde {\cal T}}}


The application of random matrix theory to problems 
in physics can be traced
back to the studies of Wigner \cite{w1,w2} in
relation to the energy spectra of complex nuclei. (For a recent survey of
the current state of the subject we refer to \cite{fsv}.) The mathematical
formalism was largely provided by Dyson \cite{d}, who 
determined that the energy level spacing distribution displays universal
behaviour depending only on
the symmetry of the Hamiltonian. (This is in contrast to the energy
distribution, i.e. density of states, which is generically
non-universal.) 

In this Letter we study the energy level spacing distribution for two
models of coupled single-mode Bose-Einstein condensates. Both of these 
models are {\it integrable} (i.e., there exists a set of mutually
commuting operators, which includes the Hamiltonian, the number of which 
is equal to the number of degrees of
freedom in the system). These models are also integrable in the sense of
the Quantum Inverse Scattering Method (QISM) \cite{kbi} 
based on the Yang-Baxter
equation, as shown in \cite{zlgm}. 
We find that the energy level spacing distribution follows the
universal Poisson form $P(s)=\exp(-s)$, where $s$ is the dimensionless   
energy gap parameter.    
Such a distribution indicates  a lack of correlation between the energy
levels, leading to random clustering. 
It was argued by Berry
and Tabor \cite{bt}, based on a semi-classical approach,
that a Poisson distribution holds 
for integrable systems 
when the number of degrees of freedom is greater
than one.  
This result has been supported by many numerical studies on
one-dimensional lattice models with many degrees of freedom,
such as the Heisenberg model, the
Hubbard model and the $t-J$ model
(at supersymmetric coupling) \cite{pzbmm,ha,am}. 
The integrability of these models results from the fact that each can
be derived  through the QISM. 


Our study for models of Bose-Einstein condensates 
is motivated by two factors. 
The first stems from the fact that the models we
will analyse are integrable for a relatively large number of independent
coupling parameters. This is in contrast to one-dimensional lattice
models where integrability generally imposes severe constraints on the
coupling parameters.
The second motivation is that these  
examples are models with low numbers of degrees of freedom acting in 
Hilbert spaces with arbitrarily large dimensions. In \cite{am} it is claimed
that the reason for an integrable system to show a Poisson distribution
for the level spacing is 
{\em there exists a basis independent of the parameters in which the
Hamiltonian is diagonal since there are as many commuting operators as the
size of the Hilbert space}. Or as explained in \cite{fsza}, if there are an
infinite number of conservation laws, then each subspace defined by a
given set of quantum numbers contains a single level. 
Since the Poisson distribution occurs for the
eigenvalue level spacings in random diagonal matrices, it should therefore
apply to integrable systems. 
However, such an argument relies on there
existing a large number of constants of the motion comparable to the
dimension of the Hilbert space. The 
findings we present here surprisingly indicate that that the result is 
true even for systems with a small number of degrees of
freedom, and does not rely on a semi-classical limit as in \cite{bt}. 

The first model we analyse is the two-site Bose-Hubbard model which has
been widely applied for the study of two coupled Bose-Einstein
condensates 
\cite{leggett}. The Hamiltonian is   
\bea
H&=& U_{11} N_1^2+U_{12}N_1N_2+U_{22} N_2^2 +\mu_1 N_1+\mu_2 N_2\no\\
&&~~~~~ -\frac{\E_J}{2} (a_1^\dagger a_2 + a_2^\dagger a_1), 
\label{aaham} \eea
where the operators $a_i,\,a^\dagger_i,\,N_i=a^\dagger_i a_i,\,i=1,2$ are
associated to two Heisenberg algebras with relations
$$[a_i,\,a_j^\dagger]=\delta_{ij},~~~[a_i,\,a_j]=[a^\dagger_i,\,a^\dagger_j]
=0. $$
The model describes Josephson tunneling
between two condensates with tunneling strength ${\E_J}/{2}$, the
parameters $U_{ij}$ are the amplitudes for $S$-wave scattering and
$\mu_i$ are chemical potentials. 
The Hilbert space of states is given by the infinite-dimensional Fock
space
spanned by the vectors
\bea  \left|m,n\right>=(a_1^{\d})^m(a_2^{\d})^n\left|0\right>, ~~~~~~~~
m,\,n=0,1,2,....,\infty . \label{fock} \eea 
As each basis vector is uniquely determined by the quantum numbers $m,n,$
this model has two degrees of freedom. The Hamiltonian commutes with the
total
particle number $N=N_1+N_2$, so the existence of the two conserved
quantities, $H$ and $N$, shows that the model is integrable.
~~\\
\begin{table}
\centerline{
\begin{tabular}{|l|c|c|c|c|c|c|}
\hline
&$U_{11}$&$U_{22}$ & $U_{12}$&$\mu_1$&$\mu_2$&$\Omega$
\\
\hline
$\blacktriangle$& $0.01$ & $0.01$ & $-0.02$ & $0.01$  & $-0.01$  & $100$
\\
\hline
$\blacksquare$& $2.0$ & $2.0$ & $-4.0$ & $1.0$  & $-1.0$  & $1.0$
\\
\hline
\,$\bullet$& $100$ & $100$ & $-200$  & $10$ & $-10$ & $0.01$
\\
\hline
\end{tabular}
}
\caption{Choices of coupling parameters for the Hamiltonian
(\ref{aaham}) used to determine the density of states shown
in Fig. \ref{fig1}.}
\end{table}

In Ref. 
\cite{zlgm} it was demonstrated that (\ref{aaham}) can be derived through
the QISM and solved exactly using the algebraic Bethe ansatz.  
Here we will solve the model in the spirit of the co-ordinate Bethe
ansatz, which allows us to exploit the 
tridiagonal structure of the Hamiltonian. Letting
\bea &&\l|\Psi\r>=\sum_{m=0}^N
\alpha_m\l(a_1^\dagger\r)^{N-m}\l(a^\dagger_2\r)^m\l|0\r>
\label{vec} \eea
it is easily shown, by directly evaluating the action of (\ref{aaham})
on (\ref{vec}), that $\l|\Psi\r>$ is an eigenstate of (\ref{aaham})
provided the co-efficients $\alpha_m$ satisfy, for $m>0$,
the recursion relation
\bea
&&\alpha_{m+1}=
X_m
\alpha_m+\frac{m-N-1}{m+1}\alpha_{m-1} \label{rec}
\eea
such that $\alpha_1=X_0\alpha_0$, $\alpha_{N+1}=0$ and
$X_m$ is defined by  
\bea
&&\frac{\E_J(m+1)X_m}{2}\equiv U_{11}N^2+(U_{11}+U_{22}-U_{12})m^2\no \\
&&~~~~~~~+
(U_{12}-2U_{11})mN+\mu_1N+(\mu_2-\mu_1)m-E \no \eea
where $E$ is the energy eigenvalue. 
It is clear that $\alpha_m$ is a polynomial in $E$ of order $m$. The
$N+1$ roots of the equation $\alpha_{N+1}=0$ gives the complete energy
spectrum for the sector with $N$ particles.
Following \cite{rktb} we can reduce the second order relation
(\ref{rec}) to a first order relation. Letting
$$\alpha_{m+1}=\alpha_0\prod_{j=0}^m X_jY_j $$
with $Y_0=1$ and substituting into (\ref{rec}) leads us to
\bea &&Y_m=1+\frac{(m-N-1)}{(m+1)X_mX_{m-1}}\frac{1}{Y_{m-1}} \label{ym}
\eea
showing that $Y_m$ admits a continued fraction expansion. This provides
a convenient means to generate the polynomials $\alpha_m$. For a given
root $E$ of $\alpha_{N+1}=0$, this value may be substituted into the
terms $\alpha_m,\,m=1,...,N$ which permits us to compute the
corresponding
eigenstate through (\ref{vec}) (although this is not necessary for our
studies here).

\begin{figure}
\includegraphics[scale=0.32]{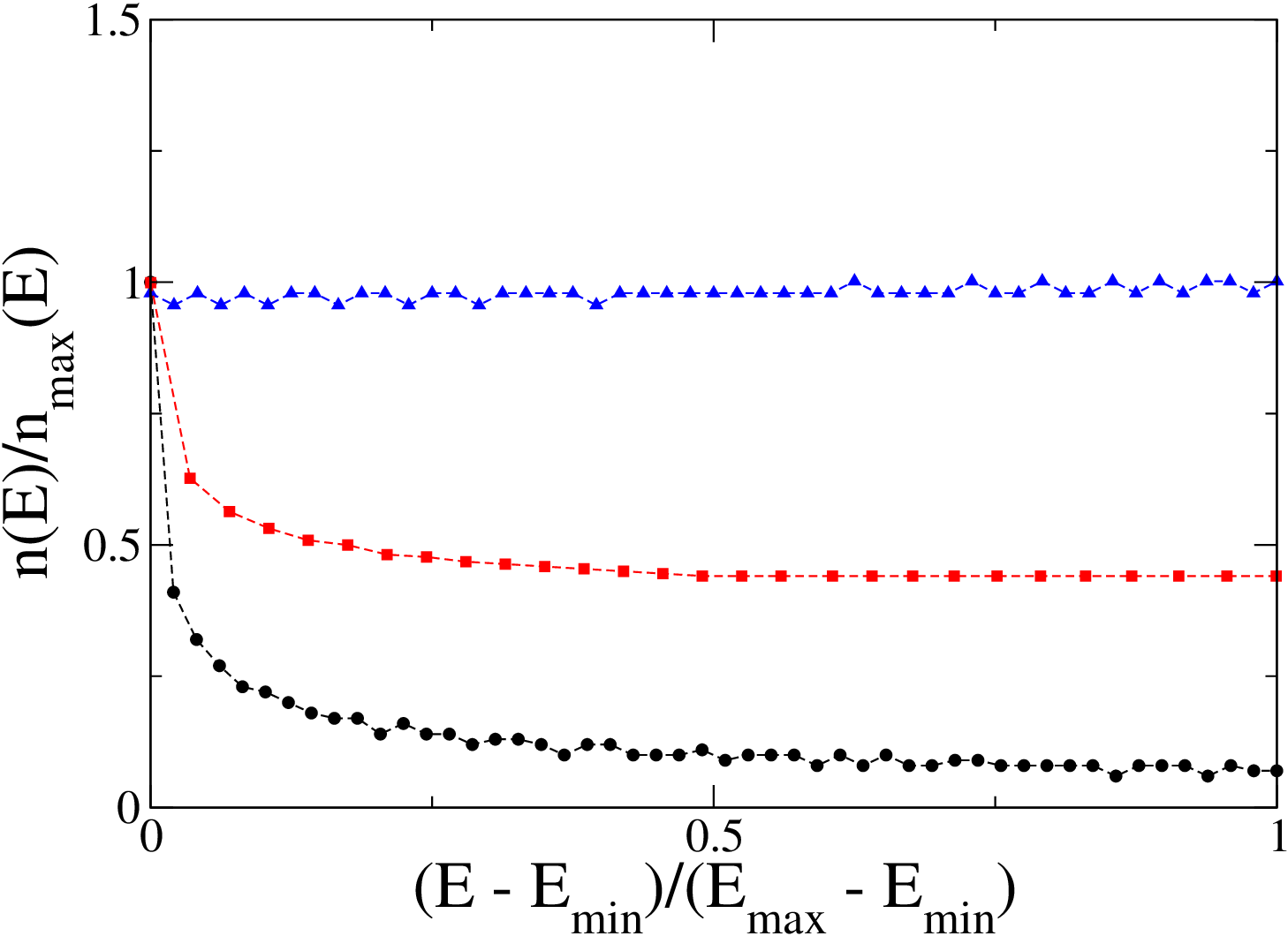}
\caption{\label{fig1}
The density of states for the two-site Bose-Hubbard model
with $1000$ particles, and interaction couplings given by 
Table I.
The approximately constant
profile in the case of the couplings ($\blacktriangle$) lying in the Rabi 
regime is reflective of the semi-classical nature of the pendulum
analogy given in \cite{leggett}. In contrast, for the couplings 
($\bullet$) lying in the Fock regime 
the pendulum analogy is not
semi-classical as indicated by a non-constant profile. The parameters
($\blacksquare$)
within the Josephson regime illustrate the crossover behaviour, 
where the density of states deviates from the 
semi-classical profile at  
low energies. }
\end{figure}

For $U\sim U_{11}\sim U_{22}\sim -U_{12}/2 $ 
it is common to divide the parameter space into three
regimes \cite{leggett}; viz. Rabi ($U/\E_J<<N^{-1}$), Josephson
($N^{-1}<<U/\E_J<<N$) and Fock ($N<<U/\E_J$).
There is a correspondence between (\ref{aaham}) in these limits
and the motion of a
pendulum
\cite{leggett}.
For both the Fock and Josephson regimes the
analogy corresponds to a pendulum with fixed length, while in the Rabi
regime the length varies. For both the Rabi and Josephson regimes the
system is semi-classical. By semi-classical we
intend that the energy per particle forms a continuum in the limit of
large particle number.
The Fock case is not semi-classical (e.g., 
there is a finite gap in the energy per particle between
the ground and first excited state; see \cite{zlmg},) and hence
the argument of \cite{bt} is not applicable. 
Fig. 1 shows the profile of the density of states $n(E)$, supporting
this picture, with the coupling parameters for the Rabi, Josephson and
Fock regimes given
in Table I. 
{
\begin{figure}
\includegraphics[scale=0.32]{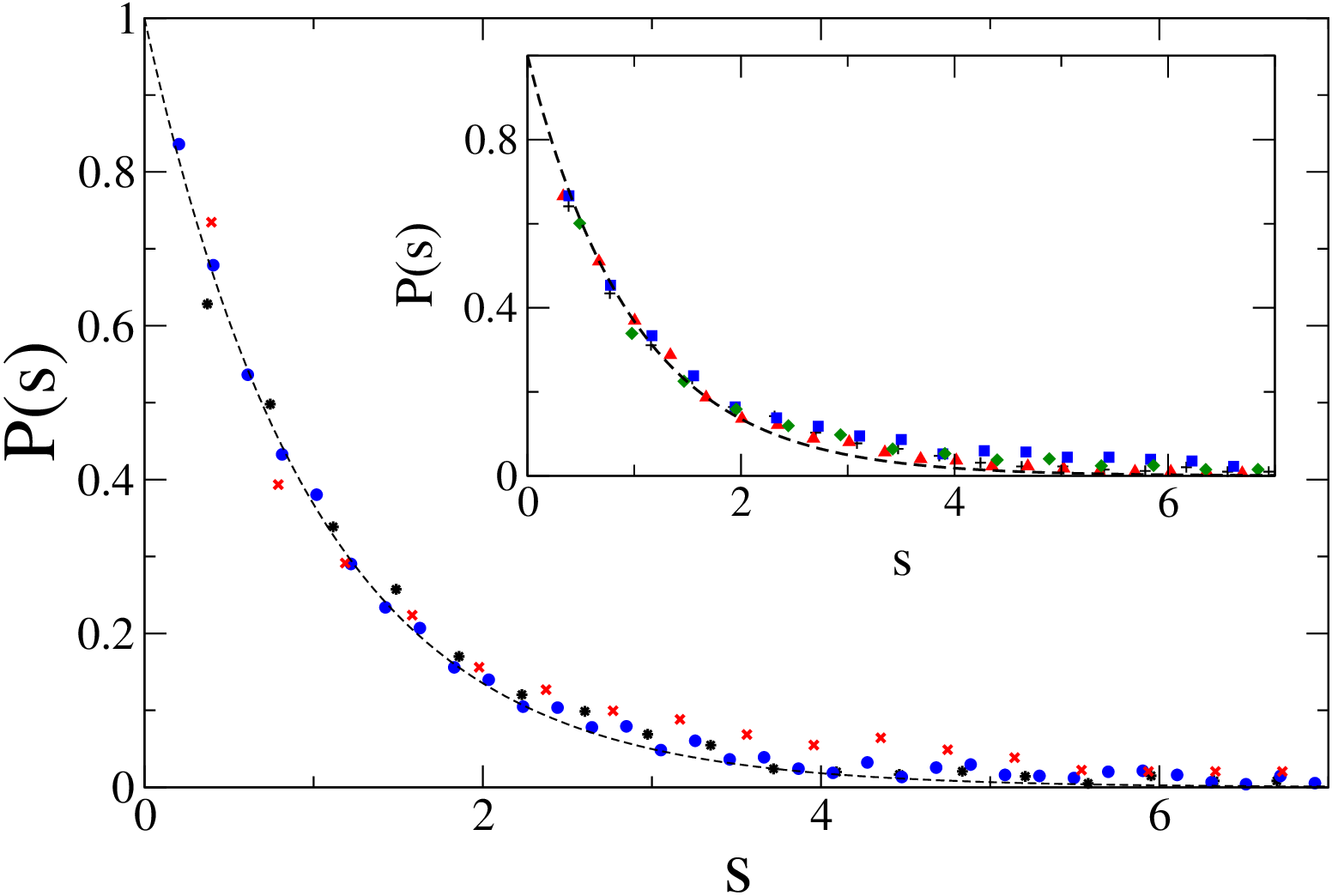}
\caption{\label{fig2}
Energy level spacing distribution for the Bose-Hubbard model (\ref{aaham})
in the sector
$360\leq N \leq 400$, where the particle numbers are increased in units
of 4.
The total number of energy levels is 4,191. 
The distribution is independent of the choice of the coupling parameters, which
are given by the first three columns of Table II. 
The inset shows the energy level spacing distribution for the remaining
couplings given in Table II. 
The results show an excellent fit to the
Poisson distribution $P(s)=\exp(-s)$, illustrated by the dashed
curve in both the figure and inset. 
\\
}
\end{figure}

To study the full level statistics of this model in the most general
context is
prohibitive because of the infinite dimensionality of the Hilbert space.
It is important to emphasize that one cannot simply restrict to a sector
of fixed particle number, as this has the effect that the restricted model
has only one degree of freedom. In this instance, we
have found that the Poisson distribution does not hold. As mentioned earlier,
the Poisson distribution is only expected to hold when the numbers of
degrees of freedom is greater than one. Therefore to investigate the
level spacing statistics, we must conduct the
analysis over a finite number of sectors  
with different particle number, in order to account for both
degrees of freedom.  
~~\\
\begin{table}
\centerline{
\begin{tabular}{|l|c|c|c|c|c|c|c|}
\hline
&$*$&$\bullet$ & $\times$&$+$&$\blacktriangle$&$\blacksquare$&
$\blacklozenge$\\
\hline
$U_{11}$& $0.01$ & $2.0$ & $100$ & $3.1$  & $-0.2$  & $66.0$  & $0.34$
\\
\hline
$U_{22}$& $0.003$ & $0.7$ & $88$ & $-0.14$  & $0.4$  & $28.0$  & $3.45$
\\
\hline
$U_{12}$& $0.0$ & $0.0$ & $0.0$  & $0.001$ & $10.0$ & $0.3$   &$0.0$
\\
\hline
$\mu_1$& $0.0$ & $0.0$  & $0.0$  & $15.0$  & $0.0$  & $3.14$  & $0.12$
\\
\hline
$\mu_2$& $0.0$ & $0.0$  & $0.0$  & $-2.0$   & $4.67$  & $143$ & $0.11$
\\
\hline
$\Omega$& $100$ & $1.0$  & $0.01$  & $0.5$  & $10.0$  & $0.24$  & $15$
\\
\hline
\end{tabular}
}
\caption{Choices of coupling parameters for the Hamiltonian
(\ref{aaham}) used in Fig. \ref{fig2}.}
\end{table}
}

We take subspaces of the Hilbert space comprised of
sectors with fixed particle number $N$ (each of dimension $N+1$),
starting at $N=360$, increasing in
steps of 4 particles,
up to $N=400$. This gives a total number of
4,191 energy levels. We calculate these energy levels for the wide range
of coupling parameters given in Table II. While the  first three choices
for the coupling parameters in Table II 
correspond
to the Rabi, Josephson and Fock regimes, the
remaining values were chosen randomly.
Note that we have deliberately
not taken $U_{11}=U_{22}$ and $\mu_1=\mu_2$ in all cases since this
corresponds to a discrete symmetry upon interchange of labels 1 and 2, 
leading to eigenvalue degeneracies which slightly complicates the analysis
of the
level spacing distribution.
~~\\
\begin{table}
\centerline{
\begin{tabular}{|l|c|c|c|c|c|c|c|c|}
\hline
&$*$&$\bullet$ & $\times$&$+$&$\blacktriangle$&$\blacksquare$&
$\blacktriangledown$&$\blacklozenge$\\
\hline
$U_{aa}$& $0.1$ & $1.618$  & $1.0$ & $1.0$ & $0.018$ & $2.0$
& $0.0$  & $22.145$  \\
\hline
$U_{bb}$& $0.1$ & $1.618$ & $1.0$ & $-3.0$ & $-0.82$ & $-19.95$
& $-1.0$  & $4.0 $ \\
\hline
$U_{cc}$& $0.1$ & $1.618$ & $1.0$ & $15.0$ & $9.55$  & $0.0$
& $12.0$ & $0.3$ \\
\hline
$U_{ab}$& $0.1$ & $1.618$ & $1.0$ & $-1.0$ & $0.23$& $10.0$
& $40.0$ & $-2.29$ \\
\hline
$U_{ac}$& $0.1$ & $1.618$ & $1.0$ & $0.5$ & $0.0$ & $0.01$
& $30.0$ & $-36.9$ \\
\hline
$U_{bc}$& $0.1$ & $1.618$ & $1.0$ & $15.0$ & $15.0$ & $-3.0$
& $-2.0$  & $0.91$ \\
\hline
$\mu_a$ & $0.0$ & $1.618$ & $0.0$ & $1.0$ & $0.4447$ & $0.0$
& $-15.0$ & $-2.0$\\
\hline
$\mu_b$ & $0.0$ & $1.618$ & $0.0$ & $-1.0$ & $-0.61$ & $-5.0$
& $-28.0$ & $5.0$ \\
\hline
$\mu_c$ & $0.0$ & $1.618$ & $0.0$ & $-5.0$ & $0.8939$ & $1.3$
& $-4.0$ & $10.34$ \\
\hline
$\Omega$& $100.0$ & $1.618$ & $0.001$ & $10.0$ & $-8.0$ & $0.1$
& $127$  & $13.7$ \\
\hline
\end{tabular}
}
\caption{Choices of coupling parameters for the Hamiltonian
(\ref{abcham}) used in Fig. \ref{fig3} and Fig. \ref{fig4}.}
\end{table}

In determining the energy level spacing distribution for Fig. 2 no
unfolding of the raw data was undertaken, nor required, unlike other
studies \cite{pzbmm,ha,am}. In each case the energy gaps
were normalised by the largest gap, and a histogram built using, on
average, 45 bins. In some cases there were a small number of large gaps
(of  
approximately two orders  of magnitude larger) relative
to the average gap, which were discarded. In all cases the number of
discarded data points was less than $1\%$ of the total. 
The curve
$y=\gamma \exp(-\beta s)$ was fitted to each of the data sets. 
Finally, each histogram was normalised by the factor $\gamma^{-1}$ and the
dimensionless 
gap parameter $s$ was rescaled in each case by $\beta^{-1}$. 
The data shown in Fig. 2 exhibits an excellent agreement with the
theoretical Poisson distribution.  
\begin{figure}
\includegraphics[scale=0.32]{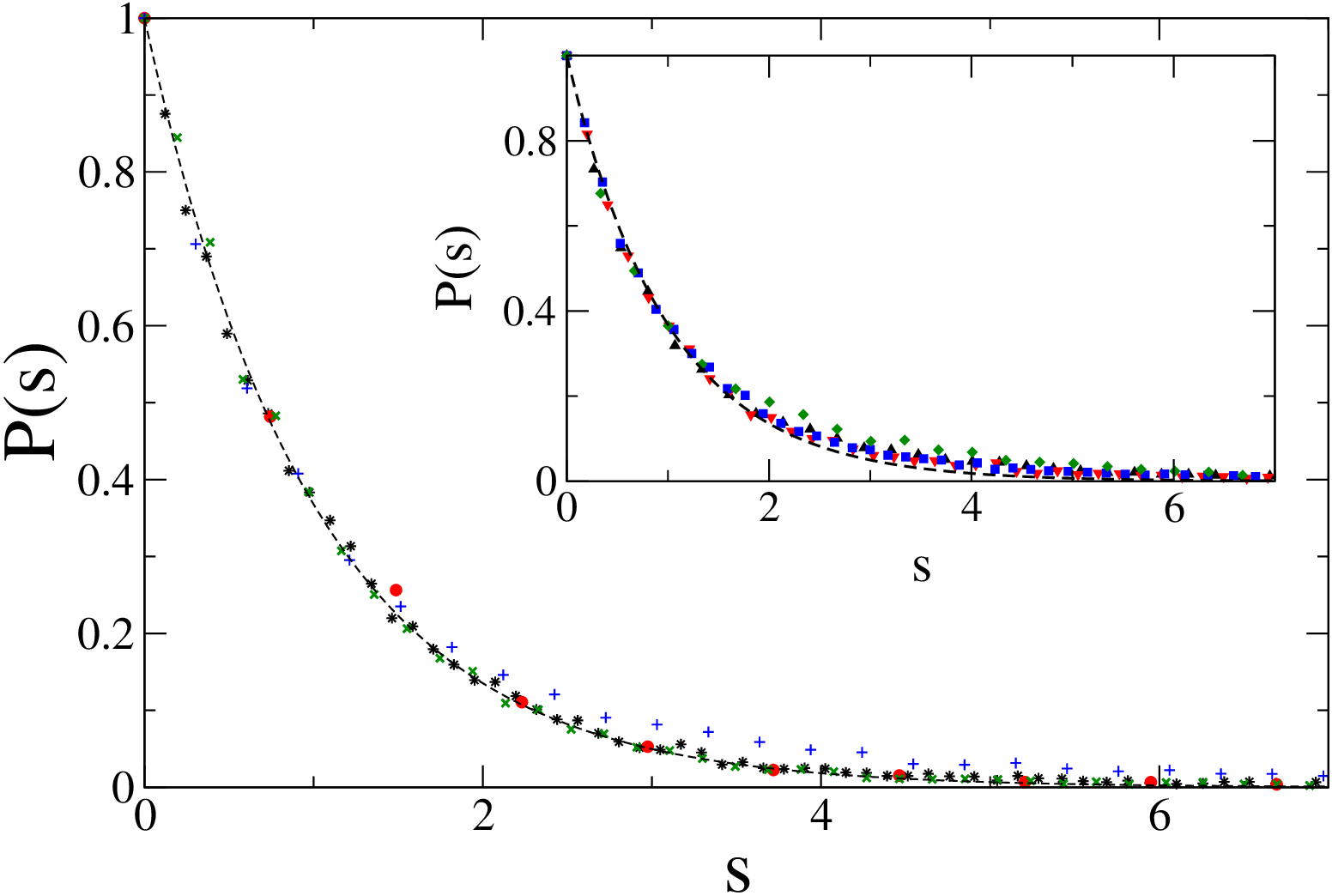}
\caption{\label{fig3}
Energy level spacing distribution for the Hamiltonian (\ref{abcham}) in
the sector with $\N=400$ and $\I\geq 0$, which has $20,301$ energy levels. 
The coupling parameters are given by the first four columns of Table III,
while the inset shows the data obtained for the remaining couplings in
Table III. The distributions have universal behaviour independent of the
choice of coupling parameters and follow the Poisson distribution 
$P(s)=\exp(-s)$, which
is illustrated by the dashed curve in both the figure and the inset.
}
\end{figure}

The model we analyse next is one for an
atomic-molecular Bose-Einstein condensate with two distinct
species of
atoms, denoted $a$ and $b$, which can combine to produce a molecule $c$
\cite{zlgm}.
It also has an interpretation as a model for second harmonic generation in
quantum optics \cite{wb} where the non-linear terms in the number
operators correspond to a Kerr non-linearity. The Hamiltonian
takes the form
\bea
H_0&=& U_{aa} N_a^2+U_{bb} N_b^2 +U_{cc} N_c^2 \no \\
&&+ U_{ab}N_aN_b+U_{ac} N_aN_c+U_{bc}N_bN_c \no\\
&&+\mu_a N_a+\mu_b N_b+\mu_c N_c
+\Omega (a^\dagger b^\dagger c + c^\dagger b a)
\label{abcham} \eea
which commutes with $\I=N_a-N_b$ and the total atomic number
$\N=N_a+N_b+2N_c$. Along with $H_0$ this establishes that the model has
three conserved integrals of motion.

The model acts on the infinite-dimensional Fock space spanned by the vectors
\bea   
&&\left| l,m,n\right>=(a^\dagger)^l(b^\dagger)^m(c^\dagger)^n\left|0\right>.
\label{fock1} \eea   
It is apparent that 
(\ref{abcham}) has three degrees of freedom, specified by the
quantum numbers $l,m,n$ in (\ref{fock1}), and is thus integrable.
For this model we can fix the total
atomic number $\N$ and make the restriction $\I\geq 0$, 
which gives a Hamiltonian with two degrees of
freedom acting in a {\em finite} Hilbert space of dimension 
$d=(\N^2+4\N+3)/8$ for $\N$ odd and $d=(\N^2+6\N+8)/8$ for $\N$ even.

The method described earlier for diagonalising (\ref{aaham}) generalises in a
straightforward way to (\ref{abcham}). Using this method we have
diagonalised (\ref{abcham}) in the sector with $\N=400$ and $\I\geq 0$ 
(this gives a total of 20,301 energy
levels) for the choice of
couplings in Table III, and determined the energy level spacing distribution in
exactly the same manner as described for (\ref{aaham}). The results
depicted in Fig. \ref{fig3} indicate a Poisson distribution 
for the level
spacings which is independent of the choice of coupling parameters. 

Finally, we also examined the energy level spacings in the non-integrable
Hamiltonian 
\bea &&\H=H_0+H_1 \label{nonint} \eea  
where $H_1=a^\dagger b+ b^\dagger a$. Note that the presence of the
term $H_1$ means that $\H$ commutes with $\N$, but does commute
with $\I$. Each sector with fixed $\N$ has dimension $d=(\N^2+4\N+3)/4$
for $\N$ odd and $d=(\N^2+4\N+4)/4$ for $\N$ even. The results shown in
Fig. 4 indicate that in this non-integrable case the level spacing
distribution no longer fits the Poisson distribution, 
but lies closer to the Wigner surmise 
\bea &&P(s)=\frac{\pi s}{2}\exp\left(\frac{-\pi s^2}{4}\right),\label{ws}\eea   
which is the distribution for the Gaussian Orthogonal Ensemble (GOE) as
expected for non-integrable 
time-reversal invariant Hamiltonians. A distinguishing
feature of the GOE is level repulsion, which was well illustrated by
Dyson \cite{d} who showed an analogy between the distributions for the
level spacings of the 
GOE and a system of unit charges, with Coulomb repulsion, confined to
the unit circle in two dimensions. 
~~\\
\begin{figure}
\includegraphics[scale=0.32]{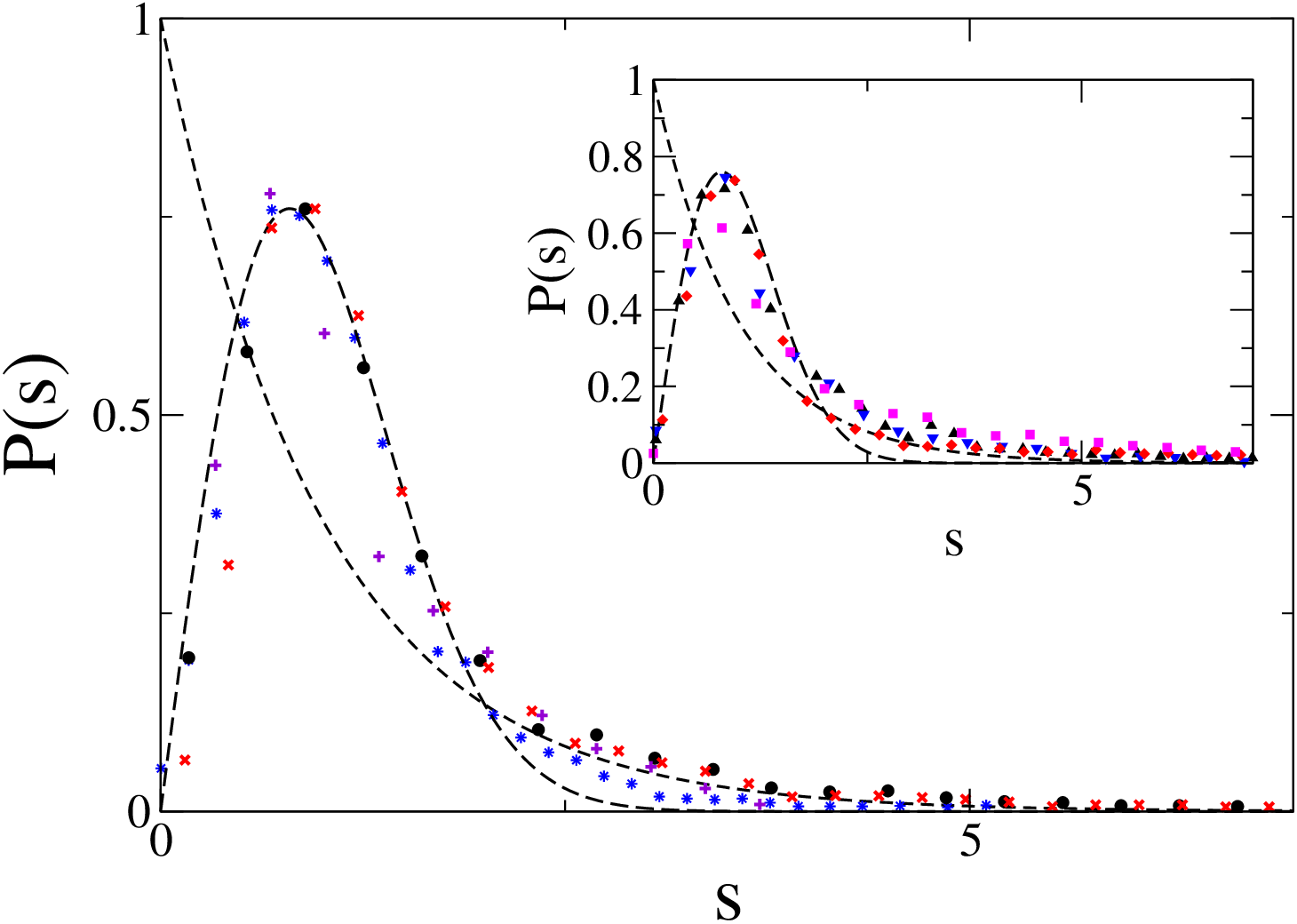}%
\caption{\label{fig4}
Energy level spacing distribution for the Hamiltonian (\ref{nonint}) in
the sector $50\leq \N \leq 100$, where the particle numbers are
increased
in units of 10. This gives a total of 9,331 energy levels.
The choice of coupling parameters is given by the first
four columns of Table II, while the inset shows the level spacing
distribution for the remaining coupling parameters. The results show a
good fit to the Wigner surmise (\ref{ws}) for a GOE distribution,
which is
indicated by the long-dashed curve (the short-dashed curve shows the
Poisson distribution).
}
\end{figure}

In conclusion, we have shown that two integrable models for systems of
Bose-Einstein condensates exhibit universal Poissonian behaviour for the
distribution of energy level spacings independent of the coupling
parameters of the Hamiltonians. Both models provide examples with a low
number of degrees of freedom acting in an arbitrarily large dimensional 
Hilbert space of states. To our knowledge this is the first study of
this type and 
complements previous studies \cite{pzbmm,ha,am} of level
spacing distributions in integrable one-dimensional lattice models.  
For lattice models the number of degrees
of freedom increases with increasing lattice length, and hence
increasing dimensionality of the Hilbert space. Our results support the
view that all integrable quantum systems show a Poisson
distribution for the energy level spacings, indicative of 
clustering due to randomness, provided the number of
degrees of freedom is greater than one \cite{bt}. Our analysis also shows
that this result does not
rely on the existence of a semi-classical limit, as it holds for cases
where the non-linear interactions dominate, such as for the Fock regime
of (\ref{aaham}). We have also studied a non-integrable example and
shown that for this case the Poisson distribution no longer holds, and
level repulsion is displayed. 
However, it is important to stress that the converse is not true: 
examples of quantum systems not exhibiting level repulsion, which are
nonetheless classically non-integrable, have been shown to exist in \cite{bss}.
(We thank Peter L\'evay for bringing this to our attention. See also
\cite{kd}.)  

This work was supported by the Australian Research Council. We thank
M.V. Berry and G.J. Milburn for helpful comments. S.R. Dahmen
thanks The University of Queensland for generous hospitality. 

\end{document}